# Towards ML/AI-based Prediction of Mobile Service Usage in Next-Generation Networks

Tarik Taleb[1,2,3], Abdelquoddouss Laghrissi[2], and Djamel Eddine Bensalem[1]
[1] Aalto University, Espoo, Finland
[2] Oulu University, Oulu, Finland
[3] Sejong University, Seoul, Korea
Emails: firstname.lastname@aalto.fi

*Abstract*—The adoption of machine learning techniques in next-generation networks has increasingly attracted the attention of the research community. This is to provide adaptive learning and decision-making approaches to meet the requirements of different verticals, and to guarantee the appropriate performance requirements in complex mobility scenarios. In this perspective, the characterization of mobile service usage represents a fundamental step.

In this vein, this paper highlights the new features and capabilities offered by the "Network Slice Planner" (NSP) in its second version [12]. It also proposes a method combining both supervised and unsupervised learning techniques to analyze the behavior of a mass of mobile users in terms of service consumption. We exploit the data provided by the NSP v2 to conduct our analysis. Furthermore, we provide an evaluation of both the accuracy of the predictor and the performance of the underlying MEC infrastructure.

## I. INTRODUCTION

Artificial Intelligence (AI) techniques are considered key technologies, able to provide precious insights by analyzing the huge amount of data generated by network devices, mobile applications, and even user behavior. Furthermore, the analysis of user mobility, as well as the characterization of mobile application usage, are getting crucial to provide useful information for service provisioning and management, thus accommodating the mobile users' expectations in densely populated areas.

To optimize the delivery of mobile services, it is important to understand where, when, and what mobile service users launch over time. With such per-service understanding of mobile traffic, supported by adequate models and predictors, the mobile traffic demand can be anticipated and both mobile services and their delivery networks can be accordingly customized. Such study of spatio-temporal characterization of mobile service consumption can also help in making a meaningful grouping of users (e.g., grouping users that move within the same neighborhood and launch the same set of services at nearly the same time). This shall help, in turn, to develop a better understanding of the aggregated behaviors of mobile users at specific locations.

One of the most common challenges for such studies is the availability of the dataset. Indeed, the data needed to characterize the users' behaviors are critical, yet not available due to concerns relevant to the privacy of users. Therefore, to cope with this challenge, in this paper, we resort to simulation. A new version of the NSP Simulator, introduced in Section IV, is used to acquire the needed data. Appropriate AI techniques are then selected to achieve accurate mobile service characterization, and an extended performance evaluation is carried out to evaluate the robustness and accuracy of the proposed techniques.

The remainder of this paper is organized as follows. Section II introduces some related research work. After a brief introduction to Machine Learning in Section III, the Network Slice Planner v2 is described in Section IV. In Section V, the method proposed to predict the usage behavior as well as the approach to handle the prediction are explained. Section VI discusses the obtained results, comparing scenarios with and without the envisioned prediction method. Finally, the paper concludes in Section VII.

## II. RELATED WORK

Machine learning techniques are deemed highly important to further optimize next generation networks [1]. In theory, machine learning can be used for automating network operations. However, the implementation in a real-life scenario faces several limiting factors (i.e., legacy network equipment, the complexity of the network, and scattered control). But with the emergence of key technologies for network Softwarization, such as Software Defined Networking (SDN) and Network Function Virtualization (NFV), networks have become more homogeneous and network control can be remotely carried out from a centralized entity. Around these technologies, authors in [2] proposed an architecture of an automated network, leveraging machine learning for a real-time balancing of resources.

In this context, Naboulsi et al [3] were able to categorize the call data records obtained from diverse operators using an unsupervised technique based on the usage characteristics of the users. This experiment allowed an accurate prediction of usage dynamics. On the other hand, the study conducted in [4] focused on finding correlations between multiple parameters such as the subscriber identifier, data consumption and access time. Such studies can help in adapting the network according to the users' needs. Nowadays, the usage of smartphones is more varied and is not only bound to simple phone calls. This has motivated several research activities to use a monitoring application to collect data related to the users' behaviors and their patterns of service usage directly from smartphones [5].

Multi-Access Edge Computing (MEC) is an emerging technology that aims to reduce end to end latency by providing a cloud infrastructure at the edge of the network and hosting services nearby end-users. When efficiently exploited, such key technology combined with artificial intelligence can help to improve the overall network performance [6, 7]. In this vein, the authors in [8] proposed applying a reinforcement learning approach in a MEC-based environment to achieve a dynamic resource allocation ensuring good performance. Such analysis could help in increasing both the Quality of Experience (QoE) and Quality of Service (QoS).

## III. MACHINE LEARNING

Machine learning aims to extract knowledge from data. This process of learning is performed through algorithms. A variety of algorithms exist, depending on the nature of the problem, and some are more suited than others. Machine learning can be divided into three main categories, each category has its own way to handle the learning [10]:

1) **Supervised Learning:** This category applies to labeled data, where the algorithm will create models that take unlabeled data instances as input and map them to their corresponding label. This approach is mostly used for resolving classification problems.
2) **Unsupervised Learning:** This kind of learning takes unlabeled data, tries to search for predominant patterns, and regroups them accordingly. This category is appropriate for clustering problems.
3) **Reinforcement Learning:** This method relies on observations to learn from the environment in an iterative fashion, then exploits the acquired knowledge to adopt the correct course of actions and reach the goal state.

Most of the machine learning algorithms belong to either one or more of the aforementioned categories. For instance, Support Vector Machine (SVM), Decision Trees and K-Nearest Neighbor (KNN) are mostly applied when treating a supervised learning problem; K-Means is a popular algorithm for resolving unsupervised learning problems. Regarding Reinforcement Learning, Q-Learning tends to be widely used among the learning policies.

## IV. NETWORK SLICE PLANNER V2

The ability to anticipate the demand for network resources requires a deep understanding and analysis of the behavior of users and that is in terms of both users' mobility and their patterns of mobile service consumption over space and time. To that end, information such as location, service usage and data consumption are crucial. This set of information could be collected either from a mobile operator or by using a monitoring application installed on the users' terminals (e.g., smartphones). Due to the critical nature of such data (i.e., since it is directly related to the privacy of users, mobile operators and users are reluctant to share data), it has become more apparent and useful to rely on realistic simulation solutions.

Indeed, even though anonymizing the data, by replacing subscribers' unique identifiers, could seem a viable solution to address this issue, the privacy could still be breached. This was portrayed in [11] where the authors conducted an analysis on the locations visited by users and were able to identify individual users according to their frequency of movement. It was presumed that the most frequently visited places could represent either home or work places.

To cope with the unavailability of such data, NSP mimics as much as possible real-life use-cases, by defining a spatio-temporal modeling of mobile service usage over a particular geographical area and in real time[1]. The generation of services is based on models that characterize the behavior of real users in consuming different types of services such as video streaming, social networks, and Internet of Things services (IoT). The mathematical background behind each service-type within the service consumption module of NSP is given in [12], and its consequent compound traffic is validated in [9]. Many virtual network function placement algorithms exploited the traces generated from NSP such as in [16]. NSP was then extended for the abstraction of the LTE workload generation, and the performance modelling of the whole LTE network [9].

In NSP, users are simulated with different mobility profiles (i.e., walking, biking, or driving) that are enhanced relying on Google APIs, and several patterns in mobile service consumption. The first version implements multiple services such as video streaming, social networks and instant messaging [12]. The new version is more aligned with 5G use-cases, as it incorporates services related to Unmanned Aerial Vehicles (UAVs) and IoT. It also simulates the cloud resource usage in terms of RAM, CPU, and storage, as well as networking-relevant events (e.g., handoff operations, Tracking Area Updates (TAU), migration, and offloading operations). Leveraging LENA NS3[2], NSP provides detailed information and statistics on Key Performance Indicators (KPIs) (e.g., temporal variation of PHY Layer KPIs such as RSRP (Reference Signal Receive Power) and SINR (Signal-to-Interference-plus-Noise Ratio) reported by User Equipments (UEs), temporal PDCP (Packet Data Convergence Protocol) Layer KPIs such as the average PDU size and delay) and events (e.g., detailed logs related to the signaling messages generated during a handoff operation such as UE's position, time, cell ID, UE's IMSI, and average SINR).

### A. Added Services

In this sub-section, we highlight the new services within NSP v2, namely UAV and IoT services.

1) **UAV Services:** UAV-related services are implemented as a drone delivery service and a drone transportation service. NSP uses Software in the Loop (SITL) with Ardupilot to simulate realistic drone behavior. Packets generated, such as defined in the MavLink Protocol[3], are then collected in addition to the network related information. When a UE requests one of the UAV-related services, a drone is then dispatched from the drone home to the destination of the UE. Such a way enables the simulation of realistic use cases of UAV services with an accurate behavior.

---
[1]Network Slice Planner: http://mosaic-lab.org/implementations.aspx
[2]The LTE-EPC Network Simulator (LENA) project: http://iptechwiki.cttc.es/LTE-EPC_Network_Simulator_(LENA)

2) **IoT Services:** IoT related services are implemented as sensors that collect diverse data. Based on the Constrained Application Protocol (COAP), accurate data usage is calculated. Three types of IoT services were envisioned: Weather and Air Pollution Services, the data of which is harvested from Open Weather API to reflect realistic weather conditions in the simulated area; and a Parking Lots Service with sensors that track the number of places available in the parking lots of the simulated area.

### B. Multi-Access Edge Computing Simulation

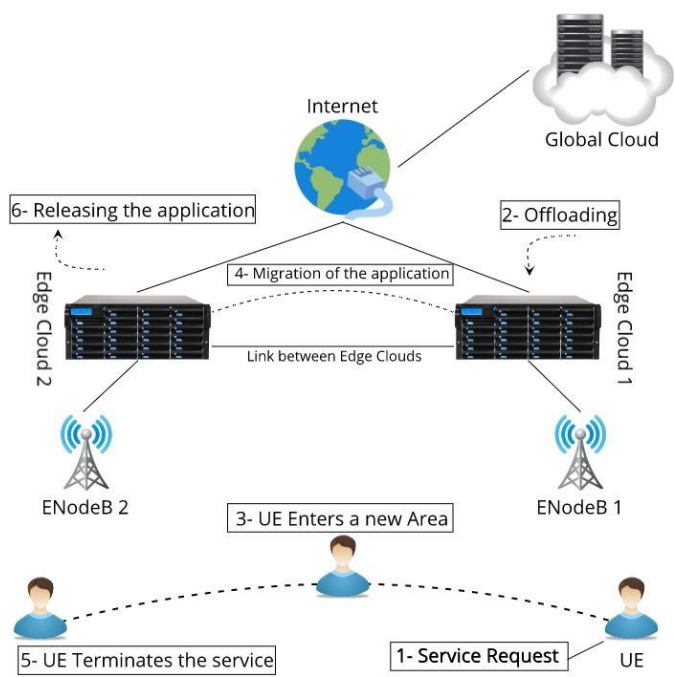

Fig. 1. Multi-Access Edge Computing Simulation Architecture.

NSP implements a Multi-Access Edge Computing Simulation module. This module aims to generate realistic data related to cloud infrastructure resources usage, and that is in terms of CPU, RAM and storage capacity. Several important events were considered (e.g., offloading and migration). The general architecture of the simulator is shown in Figure 1.

The MEC module is event-driven, where throughout the simulation and depending on the context, different events can be triggered and then processed. We defined several types of events, such as Offloading Request, Offloading Success, Offloading Failure, Migration, Migration Success, Migration Failure, Migration Aborted, and Release. The detailed mechanism of this module is given as follows.

First, when a UE starts using a service, an Offloading Request Event is triggered. Depending on the availability of the resources in the Edge Cloud (EC), an application is then transferred from the Global Cloud of the service provider to the nearest EC. It shall be noted that the term "application" refers to any kind of tasks that consumes resources in the MEC infrastructure.

Since each EC contains a fixed set of Virtual Machines (VMs), multiple policies were defined to select the appropriate VM that will carry the task. These policies are as follows:

1) **First-Fit:** it consists of choosing the first available VM that has enough resources.
2) **Best-Fit:** the application will be placed in a VM that contains the smallest sufficient amount of resources required in terms of CPU, RAM and Storage capacity.
3) **Random:** a random VM which has enough resources will be chosen as the one that will handle the application.

Based on the concept of Follow Me Cloud [13], migration events are triggered. Therefore, when a UE moves from one region to another, if a closer and more adequate EC is available, a migration process of the application from the previous EC to the new one is performed and the related events are then triggered. Finally, when the UE finishes using the service, a Release Event is triggered and the application will be terminated, freeing the resources.

## V. SERVICE PREDICTION

In order to predict the service consumption of users, it is important to analyze their behavior. In this perspective, we introduce an approach based on the concept of highly-dense areas. The idea is that before entering the phase of service prediction, a first step consists in determining areas that contain the largest number of users. Once these areas are identified, the service analysis will focus on them, thus allowing the system to predict the prevailing services consumed by a mass of users, rather than services consumed occasionally by a few individuals scattered over less-populated areas.

TABLE I
THE CONSIDERED FEATURES AND THEIR DESCRIPTION.

| Feature | Description |
|---|---|
| Time | The times in seconds |
| UE ID | The unique identifier of the user equipment |
| Service Name | The name of the service that is being used |
| Latitude | The latitude value of the GPS coordinate |
| Longitude | The longitude value of the GPS coordinate |
| ENodeB ID | The unique identifier of the ENodeB |
| Datarate Uplink | The throughput of the user in the uplink |
| Datarate Downlink | The throughput of the user in the downlink |
| Zone | The name of the target zone |

The advantage of this approach relies on its simplicity. From a mobile operator's point of view, being able to customize the network to improve both QoS and QoE for a group of users provides less complexity than managing users individually.

---
[3]Mavlink protocol: http://qgroundcontrol.org/mavlink/start

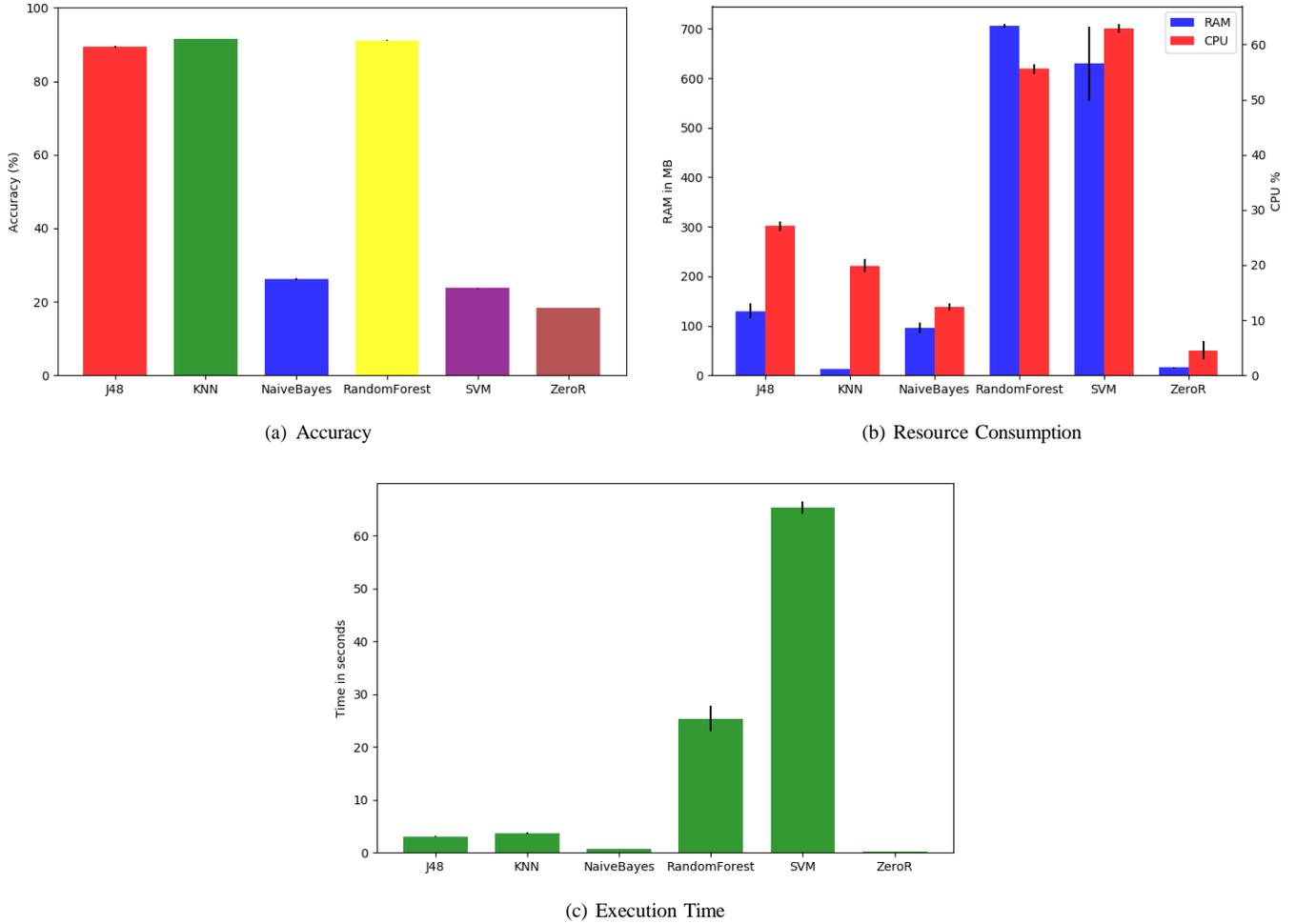

Fig. 2. Performance Benchmark of NaiveBayes, KNN, RandomForest, J48, SVM, and ZeroR

The fact that there would be always a minority of users to not benefit from the changes made in the network represents a limitation for this method.

NSP offers a variety of information. The features used in this paper are listed in Table I. As the first step for service prediction, determining high-density areas is needed. To this end, the Density-based Spatial Clustering of Applications with Noise (DBSCAN) [14] can be applied on the raw GPS coordinates of UEs. With the appropriate distance function, the results of this algorithm are clusters that represent highly dense areas. Since we use GPS coordinates, the *haversine* formula can be used as a distance function. The choice of this algorithm is mainly motivated by two reasons: *i*) this algorithm does not need to supply the number of clusters (i.e., areas) beforehand since this information is not always available in real-life scenarios, and *ii*) the ability to handle the noise. Noise points are entries that do not participate in any cluster. Indeed, those rogue points could mislead algorithms by either changing the center of the area or by mistakenly considering isolated points as a highly dense area.

With the dense areas being determined as per the previous phase, the next step consists in creating the prediction model. A classification method can be applied to the data set (i.e., generated by NSP) whereby the service name feature represents the class of the model. To evaluate different classification methods, the Waikato Environment for Knowledge Analysis (WEKA) [15] is used in this paper. This tool implements various machine learning algorithms. Its open source nature in addition to its comprehensive JAVA API (i.e., permitting its integration into the NSP simulator) are the main reasons behind choosing this tool.

## VI. EVALUATION

### A. Classification Algorithms Benchmarking

In order to select the appropriate classification algorithm for this study, a thorough analysis on several algorithms is conducted. The considered algorithms are NaiveBayes, K-Nearest Neighbors (KNN), RandomForest, J48, SVM, and ZeroR. These algorithms are selected for the wide variety of families that they represent. Many classifiers require a set of parameters. Hereunto, the default and recommended values are used.

The dataset, used to draw the results shown in Figure 2, was generated using the simulation parameters detailed in Table II and is constituted of more than $67000$ instances. Figure 2 represents the mean values and Confidence Interval (CI) of five executions.

TABLE II
SIMULATION PARAMETERS.

| Parameter | Value |
|---|---|
| Number of UEs | 500 |
| Update Meters | 100 |
| Number of eNBs per Edge Cloud | 3 |
| Number of eNBs per Tracking Area | 6 |
| Number of Edge Clouds | 10 |
| Number of Drone Homes | 5 |
| Number of Drones Per Home | 2 |
| Range | 5000M |
| Number of Weather Devices | 10 |
| Number of Air Pollution Devices | 10 |
| Number of Parking Devices | 10 |
| MIME Probability | 20% |
| Video Streaming Probability | 30% |
| Social Network Probability | 30% |
| Drone Delivery Probability | 5% |
| Drone Transportation Probability | 5% |
| IoT Services Probability | 10% |
| Simulation Time | 12H |

To evaluate the accuracy of each algorithm, the cross-validation technique was used. This technique evaluates the model by partitioning the dataset into multiple equally sized segments called folds. One sample will be used as a validation sample and the rest will be used as training samples. For our evaluation, we used a cross-validation with 10 folds. Figure 2(a) shows the accuracy obtained for each algorithm. From the result, the three classifiers KNN, RandomForest and J48 achieve the highest accuracy rates which are $91.59\%$, $91.16\%$ and $89.4\%$, respectively.

Although the accuracy may be an important parameter in the choice of a classification algorithm, the resource consumption as well as the execution time are also important factors that need to be taken into account. In this vein, a thorough analysis of the resource consumption in terms of RAM and CPU usage was conducted and the obtained results are illustrated in Figure 2(b). Clearly, RandomForest is the most resource-demanding algorithm. Indeed, RandomForest consumes roughly four times more RAM and three times more CPU than J48 and KNN. In terms of execution time, Figure 2(c) illustrates the results obtained. RandomForest still exhibits the highest values. On the other hand, KNN is slightly superior to J48 with a difference of $0.7$ s. From these results, we can conclude that KNN is the most suitable classification algorithm for our study since it offers a balance between accuracy, resource consumption and execution time.

*B. Performance Analysis*

In this section, we perform an analysis on the performance of the MEC infrastructure. By integrating the predictor into the MEC simulation module, we are able to anticipate the service usage and offload the necessary services into the EC beforehand. Thus, by comparing the performance under scenarios with and without the predictor, we can evaluate the gains brought by the anticipation of the service usage. To realize such an analysis, we use the parameters presented in Table III for the MEC simulation module. Also, several simulations were performed. The simulated scenarios share the same parameters listed in Table II and the same mobility pattern. Since the simulator generates random itineraries for UEs, the highly-dense areas change in each simulation, therefore, making the predictor obsolete since it was trained based on completely different areas. For this reason, adopting the same mobility pattern for all simulations would be a better fit to evaluate the performance. Even though they share the same parameters, the service usage of UEs changes. However, by keeping the overall probabilities of occurrence for each service the same, the general behaviors would stay nearly the same.

TABLE III
PARAMETERS OF THE MEC SIMULATION MODULE.

| Parameter | Value |
|---|---|
| Bandwidth | 1GB |
| Policy | First Fit |
| Number of VMs Per Edge Cloud | 2 |
| Host Resource | 16GB RAM, 16 Cores, 1TB Storage |
| VM Resource | 8GB RAM, 8 Cores, 500GB Storage |
| Application Resource | 1GB RAM, 2 Cores, 2GB Storage |

The results obtained by running $10$ simulations are presented in Figure 3. The results show the mean number of events occurred and their respective CI. Figure 3(a) shows the results related to the Offloading events. Two types of events are shown, namely the success of the offloading operation and its failure. The failure occurs when there are not enough resources available in the EC. We recorded $17613$ offloading requests, when using the module without the predictor. In this case, we had $8636$ successful offloading operations which represent $49\%$. On the other hand, $8977$ ($51\%$) offloading operations failed. When using the predictor, we experienced $10586$ ($60\%$) successful operations against $7027$ ($40\%$) failed operations. To conclude, using the predictor, the performance is enhanced by $11\%$.

Figure 3(b) plots the results obtained for the migration events. We had an average of $48$ Migration events without using the predictor. We then recorded three aborted migrations, 15 failed migrations, 7 successful migrations, and 23 ongoing migrations. On the other hand, by using the predictor, we had one aborted migration, 13 failed migrations, 21 successful migrations and 13 ongoing migrations. By applying the predictor, we were able to increase the number of successful migrations by $29.17\%$.

VII. CONCLUSIONS

In this paper, we introduced a new version of the Network Slice Planner (NSP), offering new services, and simulating MEC related events and resource consumption. Using the valuable information collected from the simulator, we conducted a comprehensive study on the implementation of machine learning techniques to improve both the QoS and QoE for the end-users. In this vein, a method was proposed based on using the clustering algorithm DBSCAN in order to identify highly-dense areas, thus allowing to focus on the behavior of a mass of users. By adopting supervised learning techniques through several algorithms, we were able to predict the prevailing mobile services in different areas. A detailed benchmark of diverse classifiers was conducted, and KNN was proven to be the most suitable algorithm for the envisioned study.

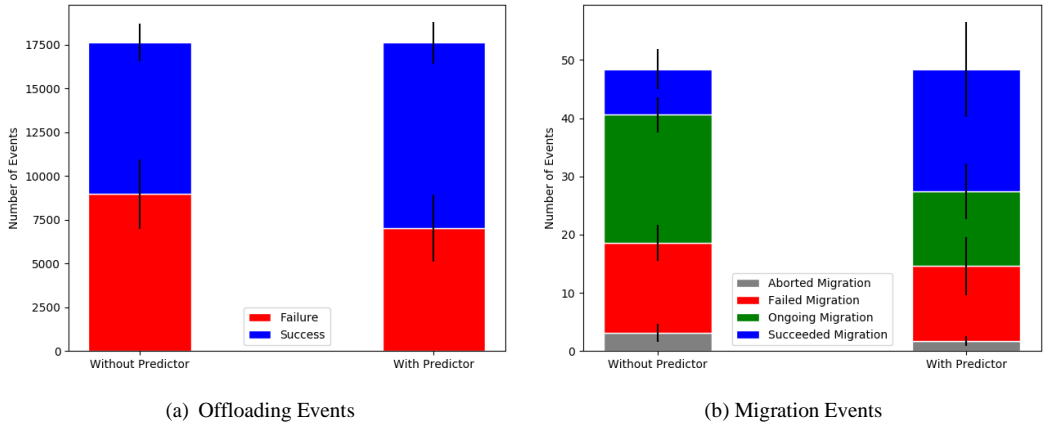

(a) Offloading Events     (b) Migration Events

Fig. 3. Performance of The MEC Infrastructure

Once the predictor is trained using the data from NSP, a thorough performance analysis of the MEC infrastructure was performed. By comparing the events of a rudimentary scenario against a scenario implementing the predictor. We noticed the gain in terms of performance. Especially, the number of successful migration and offloading operations increased by $29.17\%$ and $11\%$, respectively. Overall, this study proved the important gain that the system could get from applying machine learning techniques, ultimately ensuring a better QoS and QoE for end-users.

ACKNOWLEDGMENT

This paper was partially supported by the European Unions Horizon 2020 Research and Innovation Program through the MonB5G Project under Grant No. 871780. This work was also supported in part by the Academy of Finland 6Genesis project under Grant No. 318927 and by the Academy of Finland CSN project under Grant No. 311654.